\documentclass[twocolumn,showpacs,pre,aps,floatfix,10pt]{revtex4-1}
\ifx\pdfoutput\undefined
\usepackage{graphicx}
\else
\usepackage{epstopdf}
\usepackage{epsfig}
\usepackage{amssymb,amsmath,stmaryrd,tabularx}
\usepackage{wrapfig}
\usepackage{bm}
\fi

\usepackage[center]{subfigure}

\begin{document}

\title{Anisotropic velocity statistics of topological defects under shear flow}

\author{Luiza Angheluta$^{1,2}$, Patricio Jeraldo$^{2}$ and Nigel Goldenfeld$^{2}$}
\affiliation{$^1$Physics of Geological Processes, Department of Physics, University of Oslo, Norway\\
$^2$Department of Physics, University of Illinois at
Urbana-Champaign, Loomis Laboratory of Physics, 1110 West Green
Street, Urbana, Illinois, 61801-3080.}

\begin{abstract}

We report numerical results on the velocity statistics of topological
defects during the dynamics of phase ordering and non-relaxational
evolution assisted by an external shear flow. We propose a numerically
efficient tracking method for finding the position and velocity of
defects, and apply it to vortices in a uniform field and dislocations
in anisotropic stripe patterns. During relaxational dynamics, the
distribution function of the velocity fluctuations is characterized by
a dynamical scaling with a scaling function that has a robust algebraic
tail with an inverse cube power law. This is characteristic to defects
of codimension two, e.g. point defects in two dimensions and filaments
in three dimensions, regardless of whether the motion is isotropic (as
for vortices) or highly anisotropic (as for dislocations). However, the
anisotropic dislocation motion leads to anisotropic statistical
properties when the interaction between defects and their motion is
influenced by the presence of an external shear flow transverse to the
stripe orientation.
\end{abstract}


\pacs{61.72.CC,61.30.Jf,47.55.pb,74.40.Gh}

\maketitle
\section{Introduction}
The small-scale dynamics of interacting defects plays an important role
in the evolution of complex systems. In particular, topological defects
are a common occurrence in systems supporting a continuous symmetry
that is spontaneously broken in the process of a non-equilibrium phase
transition. One central question is how the universal properties and
scaling laws near a critical phase transition relate to the presence
and interactions of defects. The formation and evolution of topological
defects is typically formulated in the framework of the Ginzburg-Landau
theory of symmetry-breaking phase transitions, where defects are
described as phase singularities in a complex order parameter field
(rotational symmetry)~\cite{Aranson02,Pismen99}.

The equilibrium structure of isolated topological defects is deeply
rooted in their topological properties and is relatively well studied
and understood~\cite{Pismen99}. In contrast, the dynamical and
statistical properties of interacting topological defects during a
non-equilibrium phase transition are far less understood, and are the
subject of more recent systematic analyses. Numerous studies have
focused on the statistical properties of topological defect ensembles
in relation to the large-scale properties of the system. Examples range
from the quenching dynamics during phase-ordering
kinetics~\cite{BRAY94,Mazenko99}, the motion of defects in convection
patterns~\cite{bodenschatz00}, the dislocation dynamics in crystal
plasticity~\cite{Bako98,miguel2001idf} or the vortex filament motion in
quantum flows~\cite{Lathrop11}. A common characteristic of these
apparently disparate systems is that they support codimensional-two
topological defects; that is dislocations and vortices which in a
two-dimensional space (2D) become point defects, and defect filaments
or loops in a three-dimensional space (3D). One common finding is the
presence of a robust scaling law in the local velocity statistics for
these kind of defects. Recent experiments on decaying quantum
turbulence in $^{4}$He report that the velocity field $v$ induced by
quantized vortices is characterized by a $v^{-3}$ scaling, attributed
to the rare reconnection events between vortex
filaments~\cite{Lathrop08} and reproduced numerically in atomic
Bose-Einstein condensates~\cite{White10} and counterflow
turbulence~\cite{Tsubota11}. Similar velocity statistics has been
observed in a discrete dislocation dynamics model of crystal
plasticity~\cite{Weygand10} and in experiments on thermal convection in
an inclined fluid layer~\cite{Daniels03}.

Theoretically, the asymptotic tail of the velocity probability
distribution $P(v)$ can be calculated in a statistical formulation of
random stationary configurations of point defects interacting through a
logarithmic potential in a two dimensional space
(2D)~\cite{Bako98,Chavanis00}. The model predicts the same tail
distribution both in neutral systems (zero net topological charge) and
systems with a single-charge distribution. An inverse cubic scaling is
consistent with the approximation of the nearest neighbour interaction
between defects uniformly distributed in space~\cite{Chavanis00}. In
theoretical studies of defect motion during phase-ordering kinetics,
the inverse cubic law is related to the annihilation events of defect
loops or between point defects with opposite topological
charges~\cite{Bray97,Mazenko97,Mazenko99}. The coarsening during
phase ordering is reflected in a time-dependent density of defects and their
velocity distribution $F(v,t)$, which is characterized by a dynamical scaling law
related to the growth law of the characteristic lengthscale in the
ordering kinetics~\cite{Bray97,Mazenko97} (see also~\cite{Groma11}).
For non-conservative dynamics of the order parameter, the
distribution of velocity for point defects in 2D takes the form $F(v,t)
= \langle v(t)\rangle^{-1} P(v/\langle v(t)\rangle)$, where the scaling
function is $P(x)\sim x (1+x^2)^{-2}$ and the ensemble average velocity
$\langle v(t)\rangle $ at time $t$ is related to the average distance
between defects $L(t)$ at time $t$ and scales with time as $\langle
v(t)\rangle\sim 1/L(t)\sim t^{-1/2}$~\cite{Bray97,Mazenko97}. A
different scaling exponent for the scaling function $P(x)$ is predicted
for  defect filaments in three-dimensions
(3D)~\cite{Bray97,Mazenko99}, whereas experiments~\cite{Lathrop08} and
numerics~\cite{Tsubota11,Weygand10} suggest the same scaling as for
point defects.

In contrast to  isotropic vortex dynamics, dislocation motion in
crystals, as well as in stripe patterns, is typically anisotropic when
confined to certain gliding and climbing planes. In addition,
dislocations often coexist and interact with other kinds of defects
such as disinclinations and grain boundaries, which makes it harder to
study in isolation. For this reason, phase ordering is much more
difficult to study in isotropic stripe phases and polycrystalline
phases, then in anisotropic stripes and single crystals where only
dislocations are present~\cite{Boyer02,Boyer04}.

Stripe ordering is a common pattern occuring in a diversity of systems
from the zebra patterns to sand ripples and in classical fluid
convection systems, where defects are local tears of the underlying
pattern~\cite{bodenschatz00}. Anisotropic stripes or rolls develop by
an uniaxial ordering of stripes as happens, for instance, in
electrically driven convection flows of nematic liquid crystals
(electrohydrodynamic convection)~\cite{Pesch98}, or in thermal
convection flows of isotropic fluids down an inclined
plane~\cite{Daniels02}.

During relaxational dynamics, where the motion of defects is dominated
by mutual interactions prior to annihilations, the statistics of the
velocity components keeps the same form even in the presence of
strongly anisotropic motion of defects. A numerical study of 2D phase
ordering after a quench from a disordered state described by a
non-conservative time dependent real Ginzburg-Landau model showed that
the isotropic motion of point vortices is characterized by a
statistical distribution with an inverse cubic tail in the scaling
function $P(v)$, as predicted theoretically~\cite{QianMazenko03}.
Similar statistical distributions for the climbing (motion along the
direction of the stripes) and gliding (motion across the stripes)
velocities of point dislocations have been reproduced in a numerical
study of phase ordering in anisotropic stripes in two
dimensions~\cite{Mazenko06}. This is consistent with the theoretical
understanding that the dislocation motion in anisotropic stripes can be
in fact mapped onto a Ginzburg-Landau vortex-dynamics~\cite{Pismen99}.
This also means that, to the leading order approximation, the
interactions between dislocations are expected to be similar to those
between vortices.

The statistics of defect motion during non-relaxational evolution of
the system, self-sustained or driven by an external field, is less
well-understood due to correlation effects or additional driving forces
apart from the mutual nearest neighbor interactions between defects. A
self-sustained motion of defects is obtained in convection patterns
when the mean flow due to vertical vorticity, driven by the undulations
in the normal stripes and the presence of defects, acts as a
self-induced drift in the motion of defects~\cite{Greenside88}. This
non-trivial dynamics of defects leads to a spatio-temporal chaotic
state also know as~\lq defect turbulence\rq, which was observed
experimentally in fluid convection systems~\cite{Rehberg89,Daniels02}
or diffusion-reaction systems~\cite{Beta06}, and  studied in numerous
theoretical and numerical
investigations~\cite{Greenside88,Gil90,Kaiser92,Hildebrand95,Huepe04}.
In this chaotic dynamical regime, a statistically stationary
distribution of the number of defects is maintained by the defect
annihilations and the spontaneous creations of pairs due to the phase
instability. To leading order in the approximation of well mixed and
independent defects, the distribution of the number of defects follows
the Poisson statistics with mean square fluctuations given by the mean
number of defects~\cite{Gil90}. The well mixed assumption implies that
defect pairs are being created and annihilated randomly, whereas
experiments and numerical studies suggest that more often defects
created in a pair at a given time tend to annilate with each other in
the same pair at a subsequent time~\cite{Huepe04,Beta06}, which means
that correlations between defects are important effects in their
creation/annilations dynamics and leads to a modified Poisson
statistics in their number fluctuations~\cite{Hildebrand95}. A
theoretical understanding for the effect of the self-induced mean flow
on the collective statistical properties of defect motion is still
lacking. However, experimentally measured velocity statistics during
the spatio-temporal chaotic dynamics of uniaxial stripes in inclined
layer convection are observed to be slightly anisotropic and the
exponents in the tail distribution of both climb and glide motion are
close to $-3$~\cite{Daniels03}. This is suggestive of a dynamical
regime dominated by annihilations of dislocation pairs.

In this paper, we consider a simpler setup where non-relaxational
motion is driven by an externally imposed flow such that defects are
constantly created and annihilated leading to a statistically
stationary defect dynamics. This can be attained in an anisotropic
stripe system when a shear flow is acting normal to the stripe
orientation. The role of the shear flow is different from the commonly
studied case of shear alignment of isotropic
stripes~\cite{Daniels03,An09,Raman11} or the buckling instability under
shear acting along unixial stripes orientation~\cite{Granek99}.

\begin{figure}[t]
\includegraphics[width=0.8\columnwidth]{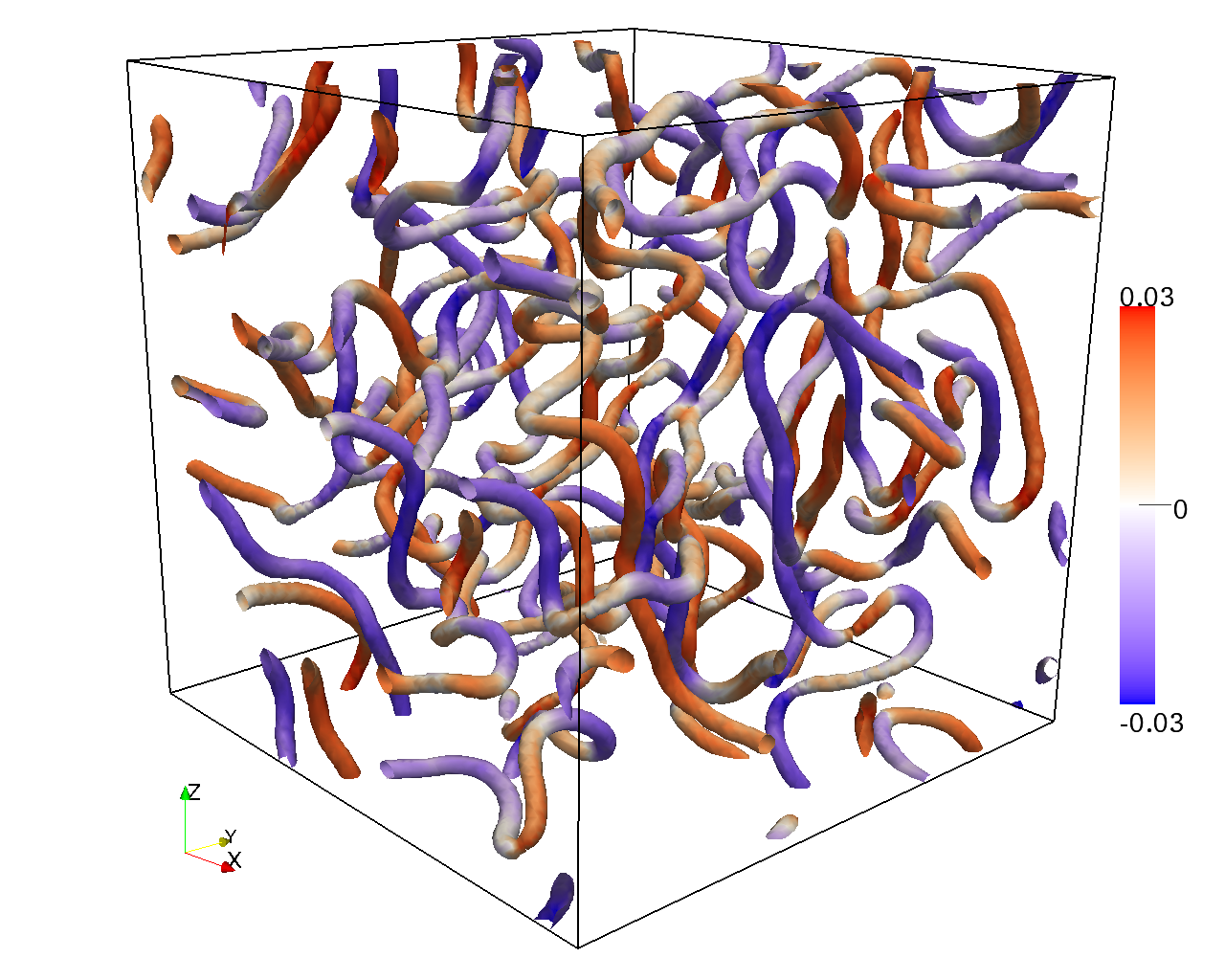}
\caption{Snapshot of a measure of the charge density vector field for a
configuration of vortex filaments in 3D simulation of phase ordering.
The figure shows the x-component $\mathcal{D}_x$ of the Jacobian
determinant defined in Sec. II. The system size in this simulation is
$128^3$. }  \label{fig:snap_3d}
\end{figure}

The purpose of this paper is threefold: i) to present an efficient
numerical method for tracking the position and velocity of topological
defects and ii) its application to study the collective motion of
dissipative vortices both in 2D and 3D and dislocations in 2D; iii) to
report on numerical results where the anisotropic motion (glide and
climb) of dislocations subjected to a simple shear flow is explicitly
manifested in the velocity distributions, even though the statistics
during phase ordering are similar to those corresponding to the
isotropic motion of vortices.

To track the position and velocity of topological defects, we
implemented a numerical method inspired by analytical treatments of
Halperin~\cite{Halperin81} and Mazenko~\cite{Mazenko97}. The method was
originally developed to locate defects in an $O(2)$-symmetric order
parameter with a Ginzburg-Landau relaxation dynamics in 2D. We show
numerically that this method works very well for Ginzburg-Landau
dynamics both in 2D and 3D and it is also suitable for tracking
dislocations in systems controlled by anisotropic Swift-Hohenberg
dynamics. Measuring the velocity statistics of vortices during
relaxational dynamics, we find a universal inverse cubic tail for
defects of the same codimension, that is point vortices in 2D and
vortex filaments in 3D. The scaling law is directly related to the
pairwise interactions between vortices prior to annihilation and
reconnection events (in 3D). Finite size core effects induce a Gaussian
cut-off to the $v^{-3}$ scaling. A similar statistical behavior is
observed in the velocity of dislocations in anisotropic stripe
patterns. Despite the fact that dislocations are dominated by their
transverse motion, and thus are highly anisotropic, the distribution of
the climb and glide velocities shows the same long tail behavior. In
the presence of an external shear flow that leads to non-relaxational
dynamics, the motion anisotropy is explicitly manifested in different
statistics of the velocity components. While the slow motion is highly
influenced by the shear flow, the high speed limit may still be dominated
by the nearest neighbor interactions.

The paper is organized as follows. Following this introduction, we
proceed in Sec.~II to discuss a method of efficiently tracking
topological defects and apply it to a collection of vortices as well as
ensembles of dislocations. Sec.~III presents numerical results on the
vortex velocity statistics in 2D and 3D simulations of phase ordering.
We discuss the statistics of dislocations during phase ordering and
non-relaxation dynamics sustained by an external shear flow in Sec.~IV.
Concluding remarks and summary are provided in Sec.~V.

\section{Defect dynamics}

Here we present a numerically efficient method for tracking
codimension-$2$ topological defects. The method is applied to
Ginzburg-Landau dynamics of vortices in 2D and 3D, as well as to
Swift-Hohenberg dynamics of dislocations in 2D anisotropic stripes. The
effect of hydrodynamic interactions in the presence of an external
shear flow is discussed in the context of dislocation dynamics.
\begin{figure}[t]
\includegraphics[width=0.48\columnwidth]{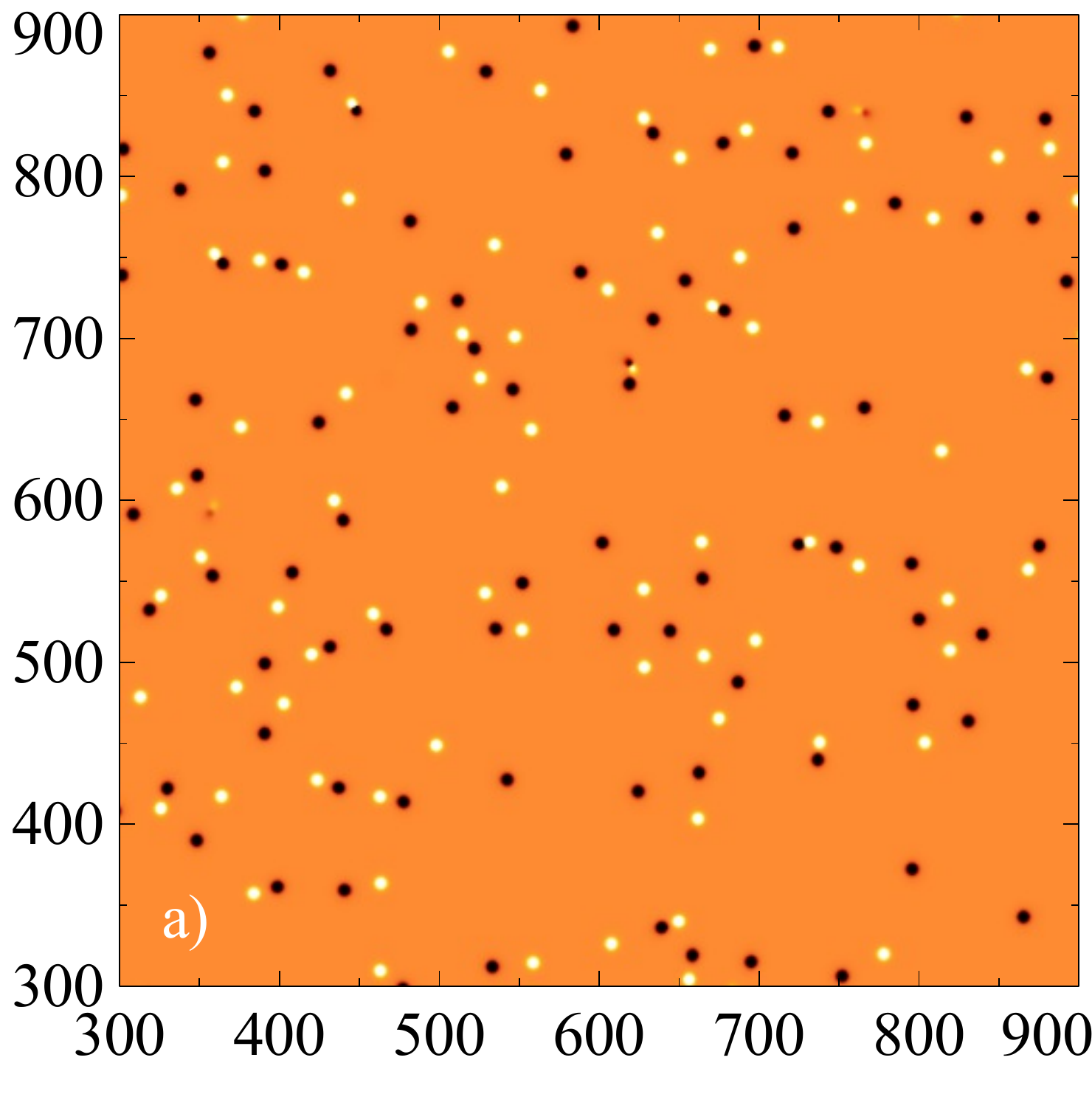}
\includegraphics[width=0.48\columnwidth]{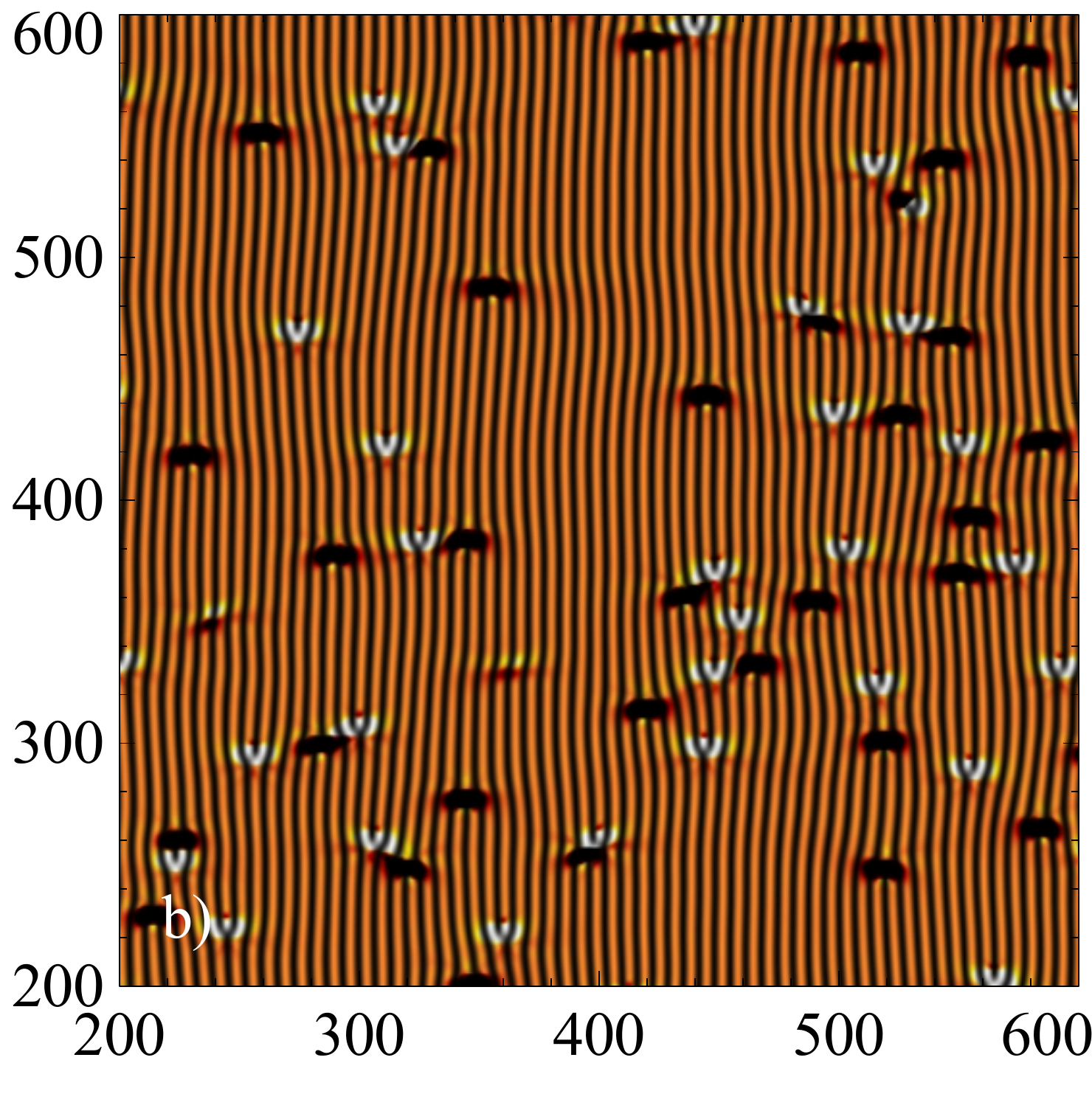}
\caption{(color online) (a) Snapshot of the charge density field
corresponding to a configuration of point vortices in 2D simulations.
In panel (b), we show the dislocations in an underlying anisotropic
stripe configuration. The lighter blobs correspond to defects that have
a positive charge, while the darker blobs are the defects of the
opposite charge. The system size in both cases is $1024^2$, while the
snapshots are drawn from a subset.}  \label{fig:snap_2d}
\end{figure}

\subsection{Locating and tracking of defects}
The identification and evolution of a large population of defects is
generally a non-trivial problem in systems described by continuum
approaches. Moving from the field variables to the discrete particle
variables is not straightforward. In most defect studies, one resorts
to various approximate methods to estimate the locations and velocity
of defects by following their
trajectories~\cite{Daniels03,QianMazenko03,Boyer04}.

In systems that can be described by an $O(2)$-symmetric order parameter
$\psi(\bm r,t)$ whose evolution depicts the ordering kinetics from an
initially disordered state to an ordered state (either isotropic and
homogeneous or a periodic pattern), one can use an elegant method based
on a transformation from the order parameter dynamics to the discrete
defect dynamics. The analytical formulation of this method was pointed
out first by Halperin~\cite{Halperin81}, and was subsequently extended
by Mazenko to determine the velocity of various topological
defects~\cite{Mazenko97,Mazenko99,Mazenko01,Mazenko03}. To our
knowledge, this method has not been previously implemented numerically.
We show that it is an efficient numerical tool used for tracking the
evolution of various types of defects.

The basic idea of this technique is that topological defects are
located at the zeroes of the complex order parameter field $\psi(\bm
r,t)$~\cite{Halperin81}. The transformation from field to particle
variables is determined by the Jacobian determinant $\mathcal{D}(\bm r)
= \vert\vert\partial\psi_n/\partial r_j\vert\vert$, where $n=1,2$
stands respectively for the real and imaginary components of the order
parameter field, i.e. $\psi = \psi_1+i\psi_2$, and $j=1,\dots d$, with
$d$ being the spatial dimension. Thus, for $d=2$, $\mathcal{D}(\bm r)$
is a scalar quantity. Its sign determines the topological charge, i.e.
$q = \mathcal{D}(\bm r)/|\mathcal{D}(\bm r)|=\pm 1$, and the charge
density is given as
\begin{equation}
\rho(\bm r,t) = \delta(\psi)\mathcal{D}(\bm r,t)=\sum_{i=1}^N
q_i\delta(\bm r-\bm r_i),
\end{equation}
for a collection of $N$ point vortices. The extension to string defects
in $d=3$ is that the Jacobian determinant becomes a vector field
$\mathcal{D}_j(\rm r)$ related to the vortex filament density
by
\begin{equation}
\rho_j(\bm r,t) = \delta(\psi(\bm r,t))\mathcal{D}_j(\bm r),
\end{equation}
where the notation for the Dirac $\delta$-function is
used~\cite{Mazenko99}.

The defect velocity $\bm v$ is determined from the property of
topological defects that their total charge is conserved (defects are
created and annihilated in pairs of opposite charge), namely
\begin{equation}\label{eq:rho}
\partial_t\rho +\nabla\cdot(\rho\bm v) =0,
\end{equation}
with the charge density $\rho(\bm r,t)$ defined above. For example, in
the case of point defects in 2D, the Jacobian determinant becomes
$\mathcal{D}=1/(2i)(\nabla_x\psi^*\nabla_y\psi-\nabla_x\psi\nabla_y\psi^*)$,
where $\psi^*$ is the complex conjugate of $\psi$. By differentiating
$\mathcal{D}$ with time, a current $\bm J^{(\dot\psi)}$ can be defined
as~\cite{Mazenko97}
\begin{equation}
\bm J^{(\dot\psi)}_\alpha = -\frac{i\epsilon_{\alpha\beta}}{2}(\dot\psi\nabla_\beta\psi^*-\dot\psi^*\nabla_\beta\psi),
\end{equation}
such that the $\mathcal{D}$-field satisfies the continuity equation
\begin{equation}\label{eq:D}
\partial_t\mathcal{D} +\nabla\cdot \bm J^{(\dot\psi)}=0.
\end{equation}
Summation over repeated indices is implied and $\epsilon_{\alpha\beta}$
is the two dimensional antisymmetric tensor,
$\epsilon_{xx}=\epsilon_{yy}=0$ and $\epsilon_{xy}=-\epsilon_{yx}=1$.
From Eqs.~(\ref{eq:rho}) and (\ref{eq:D}), the defect velocity is
determined as $\bm v = \bm J^{(\dot\psi)}/\mathcal{D}$. The defect
velocity depends on the dynamics of the order parameter through its
time derivative $\dot\psi(\bm r,t)$. Explicitly,  the velocity
components are given by
\begin{eqnarray}\label{eq:v_2d}
v_x &=& -i\frac{\dot\psi\nabla_y\psi^*-\dot\psi^*\nabla_y\psi}{2\mathcal{D}},\nonumber\\
v_y &=& i\frac{\dot\psi\nabla_x\psi^*-\dot\psi^*\nabla_x\psi}{2\mathcal{D}}
\end{eqnarray}
where $\psi^*$ is the complex conjugate of $\psi$-field and $\dot\psi$
is the time derivative of $\psi$ which determines the evolution of the
order parameter. This can be generalized to $d=3$, in which case the
velocity of vortex filaments is calculated as~\cite{Mazenko99}
\begin{equation}\label{eq:v_3d}
\bm v = \frac{\bm{\mathcal{D}}\times(\dot\psi^*\nabla\psi-\dot\psi\nabla\psi^*)}{2\mathcal{D}^2},
\end{equation}
where $\mathcal{D}^2 = \sum_{j=1}^3\mathcal{D}_j\mathcal{D}_j$ and the
velocity vector field is $\bm v = (v_x,v_y,v_z)$.

In the dilute defect density limit, it can be shown that vortex
velocity defined by Eq.~(\ref{eq:v_2d}) becomes a function of the phase
and amplitude gradients of the order parameter $\psi$ near the vortex
core~\cite{Mazenko01}. The formula is exact and applies equally well
for a high density of defects.

\subsection{Application to vortices}
Vortices are defined as the zeroes of an order parameter $\psi(\bm
x,t)$ with rotational symmetry (complex field)~\cite{Pismen99}. The
fact that the complex field vanishes at the core of a defect is
equivalent to a phase singularity, i.e. the phase of the order
parameter varies discontinuously around a closed contour surrounding
the defect. The phase $\theta$ is obtained from $\psi =
|\psi|e^{i\theta}$. The shift in phase around the contour or the
winding number, i.e. $\oint\nabla\theta\cdot d\bm l =2\pi n$, defines
the topological charge of the defect. A single vortex corresponds to a
unit of topological charge, that is $n=1$.
\begin{figure}[t]
\includegraphics[width=0.9\columnwidth]{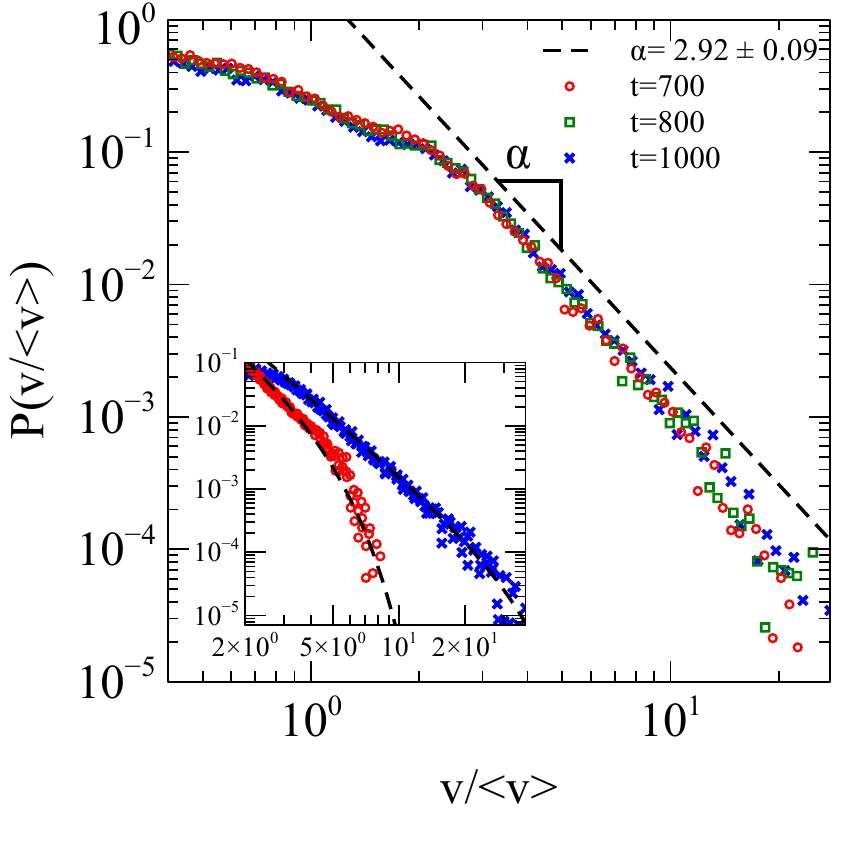}
\caption{(Color online) Collapsed probability distribution function
(scaling function) of absolute velocity of point vortices in 2D during
phase ordering. (Inset) PDF of the defect velocity for different core
sizes (open circles correspond to larger core size and  crosses
correspond to smaller core size) to show that the Gaussian cut-off
depends effectively on the vortex core size. The model parameters for
the inset figure are $A = 2.05$ (open circles), $A= 1.05$ (crosses) and
$C= 3/20(1+A)$. In the main graph, $A=1.5$. Here $v\equiv |\bf{v}|$.}
\label{fig:PDF2d}
\end{figure}

We now consider the nonconservative evolution of a $\psi({\bm r},t)$-field
described by the time dependent Ginzburg-Landau equation given by
\begin{equation}
\partial_t\psi = \nabla^2\psi + \psi
(1-|\psi|^2),
\label{eq:TDGL}
\end{equation}
which we simulate both in 2D and 3D. For computational efficiency, we
solve Eq.~(\ref{eq:TDGL}) by a cell dynamical system (CDS) algorithm,
that was originally developed for studying spinodal decomposition
dynamics~\cite{Oono1987} and extensively used to study phase ordering
of systems with continuous symmetry
\cite{Mondello1990,Mondello1992,BRAY94}.  In the Appendix we provide a
detailed description and recapitulation of the algorithm, for
completeness, and define the parameters of the simulation used below.
In particular, the depth of the quench corresponds to the parameter
$A$, and the strength of the diffusive couplings in the model are
denoted by $C$. Simulations in 2D are done on a system size of $1024^2$
cells, while in 3D we use $128^3$ cells for $dx=1$. Unless otherwise
noted, the values for the CDS parameters $A$, the depth of the quench,
and $C$, the strength of the spatial coupling, are $A= 1.5$, $C= 3/20
(1+A)$ (for 2D), and $C= 3/24 (1+A)$ (for 3D). Results were
averaged over 48 random initial conditions, unless otherwise noted.

The vortex dynamics from the Ginzburg-Landau evolution in
Eq.~(\ref{eq:TDGL}) is similar to a previous one reported in
Ref.~\cite{QianMazenko03}. Here, we use a different tracking method for
locating the defects and extend the analysis to vortex filaments in
3D.

A snapshot of the charge density field for vortex filaments in 3D
obtained using Halperin and Mazenko's method, discussed in the previous
section, is shown in Fig.~(\ref{fig:snap_3d}). The charge density field
is directly proportional to the $\mathcal{D}$-field, which is zero
everywhere except along the vortex filaments. A similar representation
is obtained for point vortices in 2D, where the charge field is
localized at the vortex core and vanishes everywhere else as shown in
Fig.~(\ref{fig:snap_2d} (a)). Since the charge density is directly
related to the $\mathcal{D}$-field, it means that the defect velocities
are meaningfully defined only at the defect positions. In the numerical
discretizations, defects are associated with small blobs (in 2D) or
thin tubes (in 3D) with a specific characteristic size that defines the
vortex core size. We define the defect regions as the locations at
which the absolute charge density is above 75\% of
the theoretical value of $|q| = 1$. The values of the
$\mathcal{D}$-field are finite within these regions and thus the
division is also finite. The velocity of the located defects is
determined from Eq.~(\ref{eq:v_3d}) for filaments in 3D, and
Eq.~(\ref{eq:v_3d}) for point vortices in 2D. The time derivative
$\dot\psi$ of the $\psi$-field is defined by the right hand side
expression in Eq.~(\ref{eq:TDGL}), namely $\dot\psi \equiv \nabla^2\psi
+ \psi (1-|\psi|^2)$.

\subsection{Application to dislocations}
Here, we focus on tracking dislocations in anisotropic stripe patterns.
A similar tracking method can be extended to locate the defects in a
crystal phase, and will be the subject of a separate study reported
elsewhere.

We consider the defects in a periodic pattern characterized by a
preferred wavenumber $\bm k$ formed by stripes.  Stripe patterns occur
in a variety of systems, typical examples being the convective rolls in
Rayleigh Benard convection of isotropic fluids~\cite{bodenschatz00} or
convective flows in the nematic liquid crystals~\cite{Zapotocky95}, in
the dynamics of diblock copolymers~\cite{christensen98}, etc.. When the
orientation of the local ordering is random, as in isotropic stripes,
we encounter both isolated defects such as dislocations or
disinclinations as well as grain boundaries. The coexistance of
different types of defects makes it difficult to analyse their
statistics. Moreover, the ordering kinetics of isotropic stripes tends
to be dominated at large times by grain boundary slow motion, which
can lead to glassy configurations~\cite{Boyer02}. By fixing the
orientation of the stripes along a preferred axis, point defects such
as dislocations can be isolated from the other types of defects.
Examples of anisotropic stripes are in electrohydrodynamic convection
of planary aligned liquid crystals~\cite{Rehberg89}, or
Rayleigh-Bernard convection of an inclined fluid
layer~\cite{Daniels03}.

\begin{figure}[t]
\includegraphics[width=0.9\columnwidth]{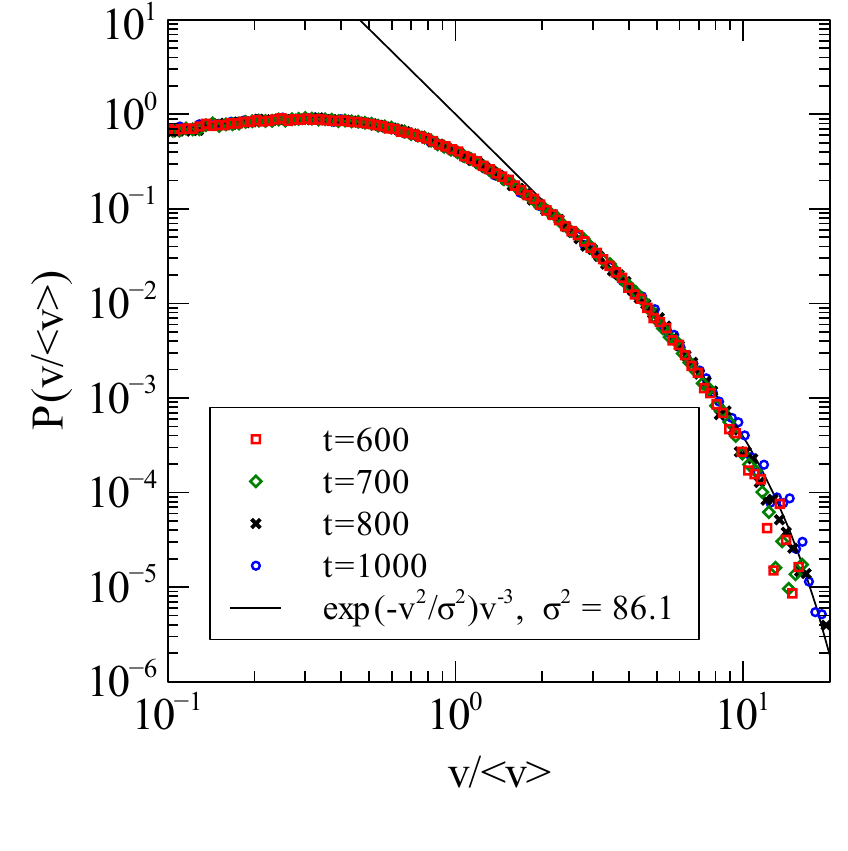}
\caption{(Color online) The scaling function of the probability
distribution function of the absolute velocity of vortex filaments in
3D during phase ordering. Here $v\equiv |\bf{v}|$.}  \label{fig:PDF3d}
\end{figure}

The statistics of dislocations in anisotropic stripes has been
discussed previously by Qian and Mazenko~\cite{Mazenko06}, where they
propose a model based on an effective Swift-Hohenberg (SH) free energy
with an additional term that accounts for the coupling to an external
field aligning the stripes along a preferred direction. The stripe
pattern in 2D is represented by a real  periodic field $u(\bm r,t)$,
which satisfies an anisotropic SH-dynamics given by
\begin{equation}\label{eq:AniSH}
\partial_t u= (1-r|u|^2)u-(1+\nabla^2)^2u-c\nabla_x^2u,
\end{equation}
where the last term is added to impose a preferred orientation of the
stripes along the vertical $y$-axis with $c>0$ being the coupling
strength to the external field. The quench depth $r>0$ is
interpreted in the context of convection patterns as the deviation from
the onset of convection, $r\approx R/R_c-1$, where $R$ is the Rayleigh
number and $R_c$ is the critical R at the onset~\cite{bodenschatz00}.
The anisotropic preferred orientation can be seen by linearizing
Eq.~(\ref{eq:AniSH}) around the mode solution $u\sim \exp(\omega t +i
k_x x+i k_y y)$ with the growth rate obtained from Eq.~(\ref{eq:AniSH})
as
\begin{equation}
\omega = (1-r)-(1-k_x^2-k_y^2)^2+c k_x^2,
\end{equation}
and imposing the condition $\omega(r;k_x,k_y) =0$ at the onset of
instability with respect to a mode of wavenumbers $k_x$ and $k_y$. The
condition is found by minimizing $\omega$ with respect to $k_x$ and
$k_y$, i.e. $\partial \omega/\partial k_x = 0$ and $\partial
\omega/\partial k_y = 0$. This leads to $k_y=0$ and $k_x =
\sqrt{1+c/2}$ for $c>0$.

We consider the amplitude formulation of Eq.~(\ref{eq:AniSH}) with an
additional contribution due to an external shear flow. We impose a
shear flow that is normal to the main orientation of the stripes to
allow for the nucleation of defects due to wavenumber shifts by shear
deformation. In order to track dislocations in a complex order
parameter field, we write the periodic field in terms of its complex envelope
field $\psi(\bm r,t)$, namely $u(\bm r) = \sqrt{r}\psi(\bm r)e^{i\bm
k\cdot\bm r}+c.c.$. Without loss of generality, we consider stripes
with the wavevector parallel to the horizontal $x$-axis, i.e. $\bm k =
(k_0,0)$. The complex $\psi$-field satisfies an amplitude equation
derived from Eq.~(\ref{eq:AniSH}) and given to the leading order in $r$
as
\begin{eqnarray}\label{eq:SH}
\partial_t \psi+\dot\gamma_x\mathcal{L}_x[\psi] = r(1-|\psi|^2)\psi-\mathcal{L}^2[\psi]-c\mathcal{L}_x^2[\psi],
\end{eqnarray}
where $\mathcal{L} \equiv (\nabla^2+2i\bm k\cdot\nabla)$ is derived
from $(1+\nabla^2)$ and $\mathcal{L}_x \equiv \nabla_x+ik_0$ comes from
the gradient $\nabla_x$. The advection term is determined by a velocity
field, which hereby is taken as a simple shear flow $\dot{\bm \gamma}=
v_{0} y \hat{\bm x}$, and the last term is added to impose a preferred
orientation of the stripes long the vertical $y$-axis.

We integrate numerically Eq.~(\ref{eq:SH}) using a $4^{th}$-order
Runge-Kutta scheme and a spherical approximation for the
gradients~\cite{Tomita91} (see Appendix) on a square domain of size
$1024 dx\times 1024 dx$. The time step is $dt= 0.05$ and the spatial
resolution is $dx = \pi/4$ so that about $8$ grid points are used to
resolve the pattern wavelength $\lambda = 2\pi/k_0$, with $k_0 = 1$.
The other parameters are set to $c=1$, $r=1$ and $v_0$ is a changing
parameter. In the absence of shear, period boundary conditions on all
sides are used. At a finite shear rate, we impose a zero flux boundary
conditions of the upper and lower boundaries and periodic conditions on
the lateral boundaries.

Dislocations are efficiently located as the zeros of the complex
envelope field $\psi$ using Mazenko's algorithm. In
Fig.~(\ref{fig:snap_2d} panel (b)), we illustrate a stripe
configuration with the location of dislocations and their topological
charge proportional to the Jacobian determinant
$||\partial\psi_n/\partial r_j||$. The velocity of dislocations is
obtained using Eq.~(\ref{eq:v_2d}) with the evolution of the order
parameter given by the right hand side of Eq.~(\ref{eq:SH}), namely
$\dot\psi\equiv
r(1-|\psi|^2)\psi-\mathcal{L}^2[\psi]-c\mathcal{L}_x^2[\psi]$.

\section{Vortex Statistics}
To determine the velocity statistics of vortices, we initiate the
system in a disordered state and follow the ordering kinetics dominated
by the initial formation and subsequent coarsening of topological
defects. At a particular time, we calculate the local defect velocities
$\bm v$ using Mazenko's method as described above. We save the absolute
values, $v = |\bm v|$, every few time iterations and run the system
from 48 random initial conditions. This way, we compute the probability
distribution function of the defect velocity at a given time, i.e.
$F(v,t)$. In the asymptotic limit $t\rightarrow\infty$ of the
coarsening dynamics, we expect scale invariance of the typical
coarsening lenghscale, i.e. $L(t)\sim t^{1/2}$ for non-conservative
dynamics (apart from logarithmic corrections in 2D). Hence the
typical velocity obtained as $1/L(t)$ scales with time as $\langle
v(t)\rangle\sim t^{-1/2}$ and corresponds to the velocity of defects in
a pair prior to annihilation and separated by a distance of the order
of $L(t)$. In the simulations, $\langle v(t)\rangle$ is the ensemble
average velocity at a particular time, and when calculated over long
times it converges to the expected asymptotic scaling. This dynamical
scaling of the mean velocity implies also a scaling with time of
$F(v,t)$. We notice that the time dependence in the PDF's can be
eliminated by rescaling the velocity variables by their ensemble
average values at a given time, i.e. $\tilde v\equiv v/\langle
v(t)\rangle$. Analytically, this corresponds to the rescaling $F(v,t) =
t^{1/2}P(vt^{1/2})$.

\begin{figure}[t]
\includegraphics[width=0.46\columnwidth]{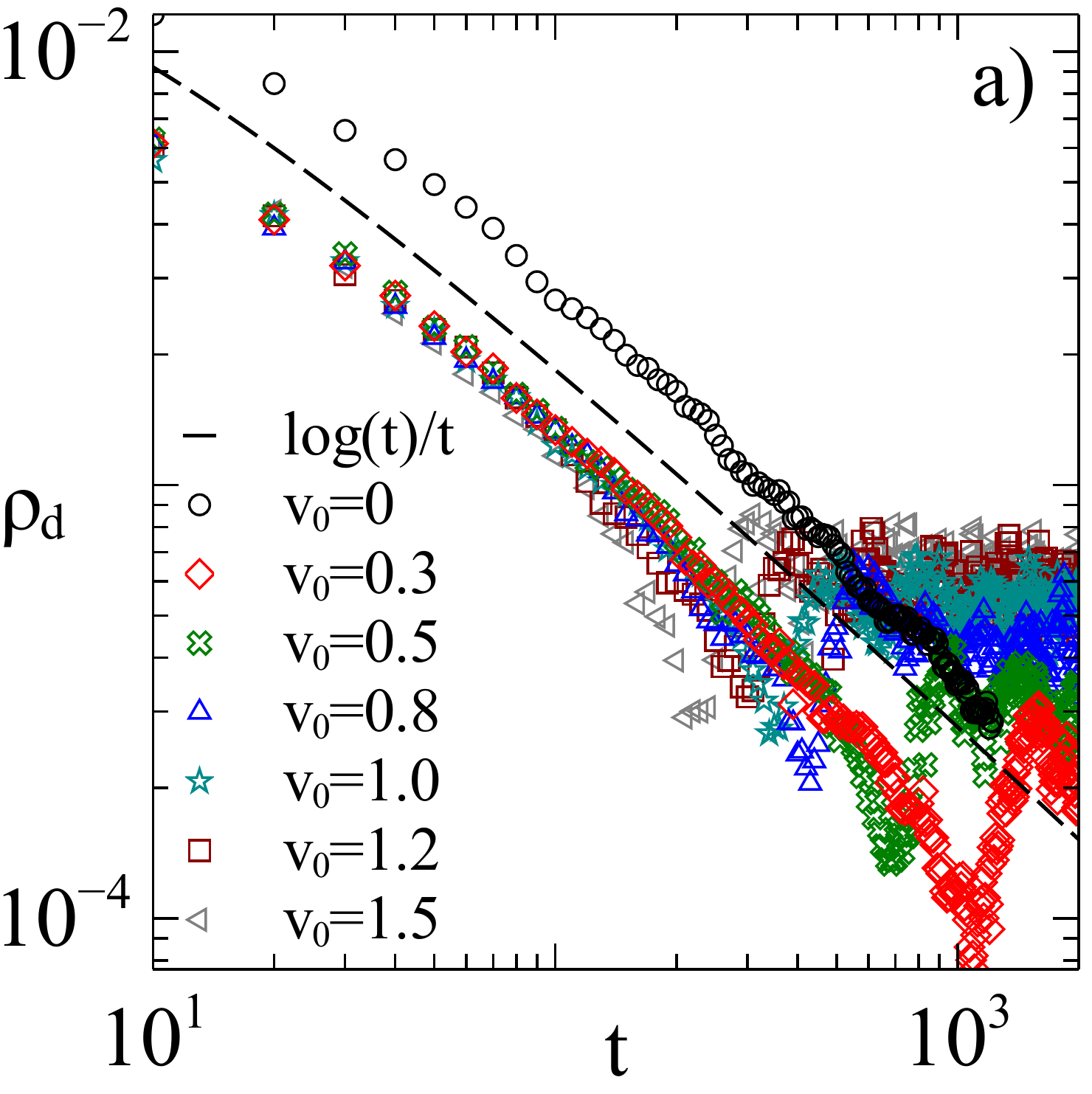}
\includegraphics[width=0.46\columnwidth]{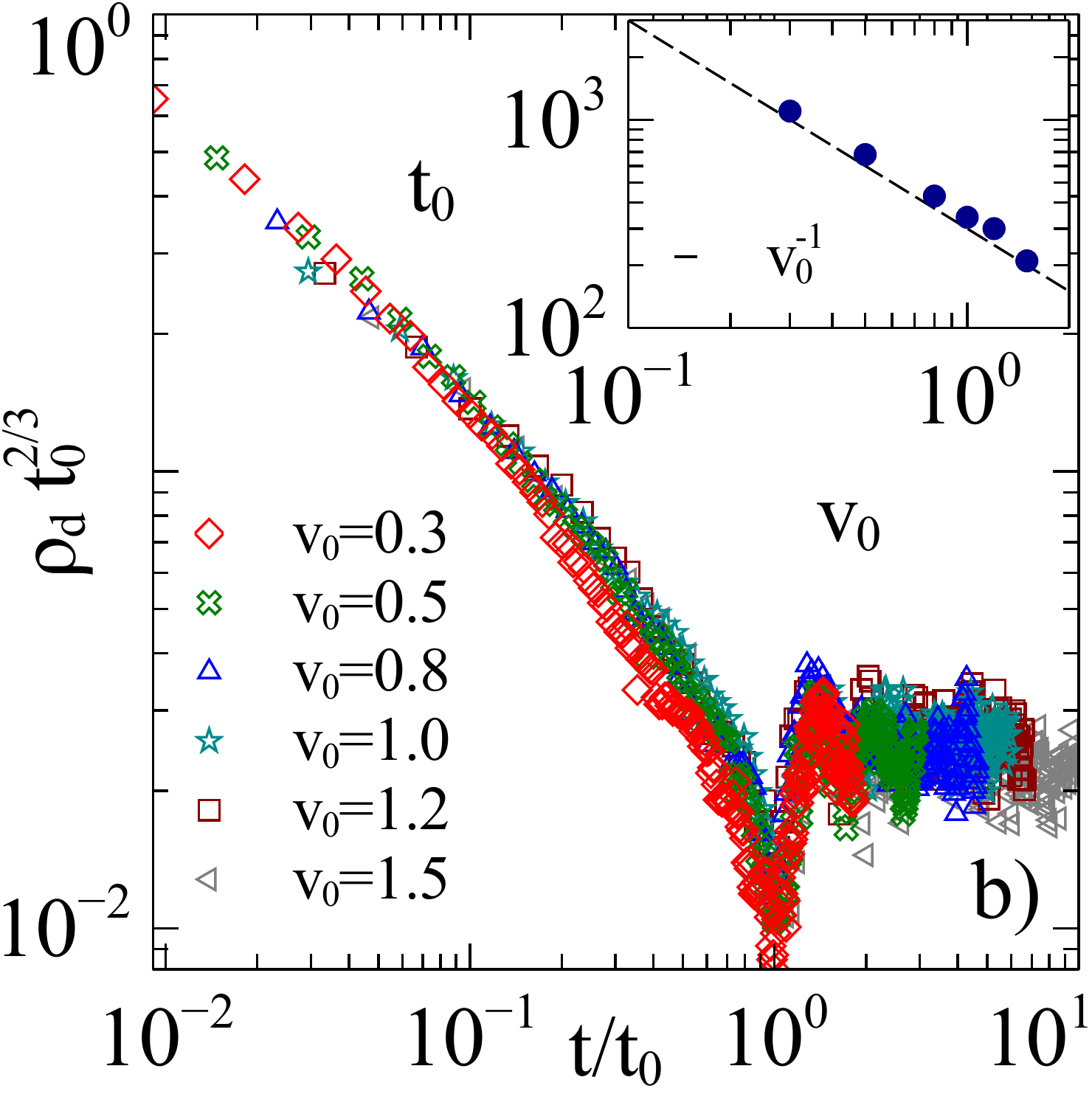}
\caption{(Color online) a) Density of defects $\rho_d$ as a function of
time $t$ for different values of the imposed shear velocity $v_0$. b)
Data collapse of the rescaled density with the mean density in the
statistical steady state $\rho_0 \sim t_0^{-2/3}$, where $t_0$ is the cross-over
time to the steady state. Inset figure shows how $t_0$ scales with
$v_0$ as $t_0\sim 1/v_0$. }  \label{fig:rho_t}
\end{figure}

The scaling function $P(x)$ of the velocity distribution is a function
of the rescaled velocity field $v/\langle v(t)\rangle$ and has a broad
tail with an inverse cubic decay. This is shown in
Fig.~(\ref{fig:PDF2d}) for 2D simulations and Fig.~(\ref{fig:PDF3d})
for 3D dynamics. The $v^{-3}$ tail corresponds to the regime of large
velocities obtained in pair interactions prior to annihilation events
or, for 3D, also reconnections events. The $-3$ scaling exponent is
determined by the logarithmic mutual interaction potential as shown in
e.g. Refs.~\cite{Bray97,Min96}. Since point defects in 2D and
filaments in 3D are both codimensional-$2$ topological defects with
the same type of interactions, we expect a similar scaling behavior.
This is consistent with other numerical studies that also show that the
tail of $P(v)$ distribution is dominated by the $v^{-3}$ scaling both
in 2D and 3D simulations~\cite{Weygand10,Tsubota11}. We provide
additional numerical evidence for this scaling regime also during phase
ordering kinetics.

\begin{figure}[t]
\includegraphics[width=0.46\columnwidth]{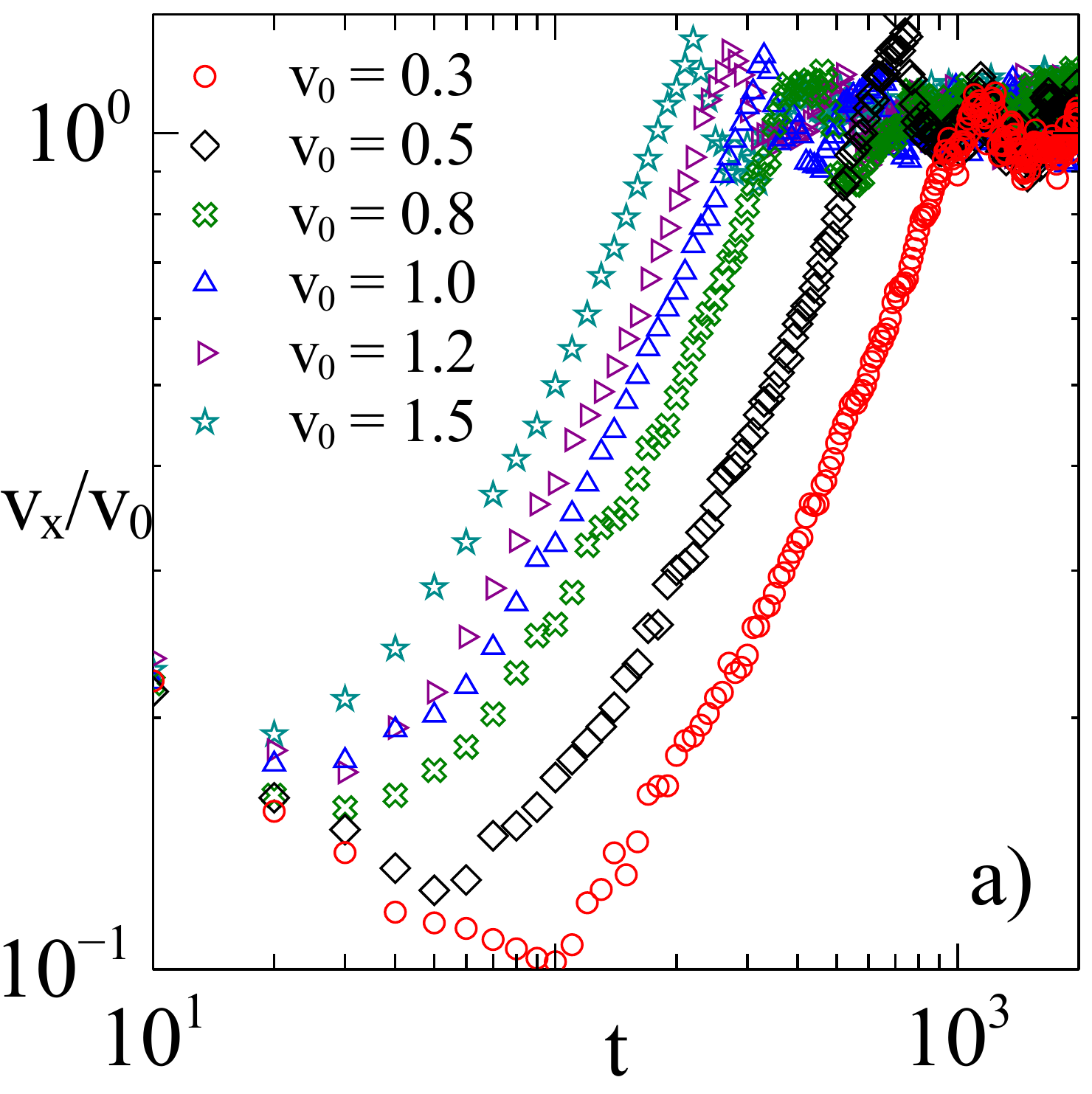}
\includegraphics[width=0.46\columnwidth]{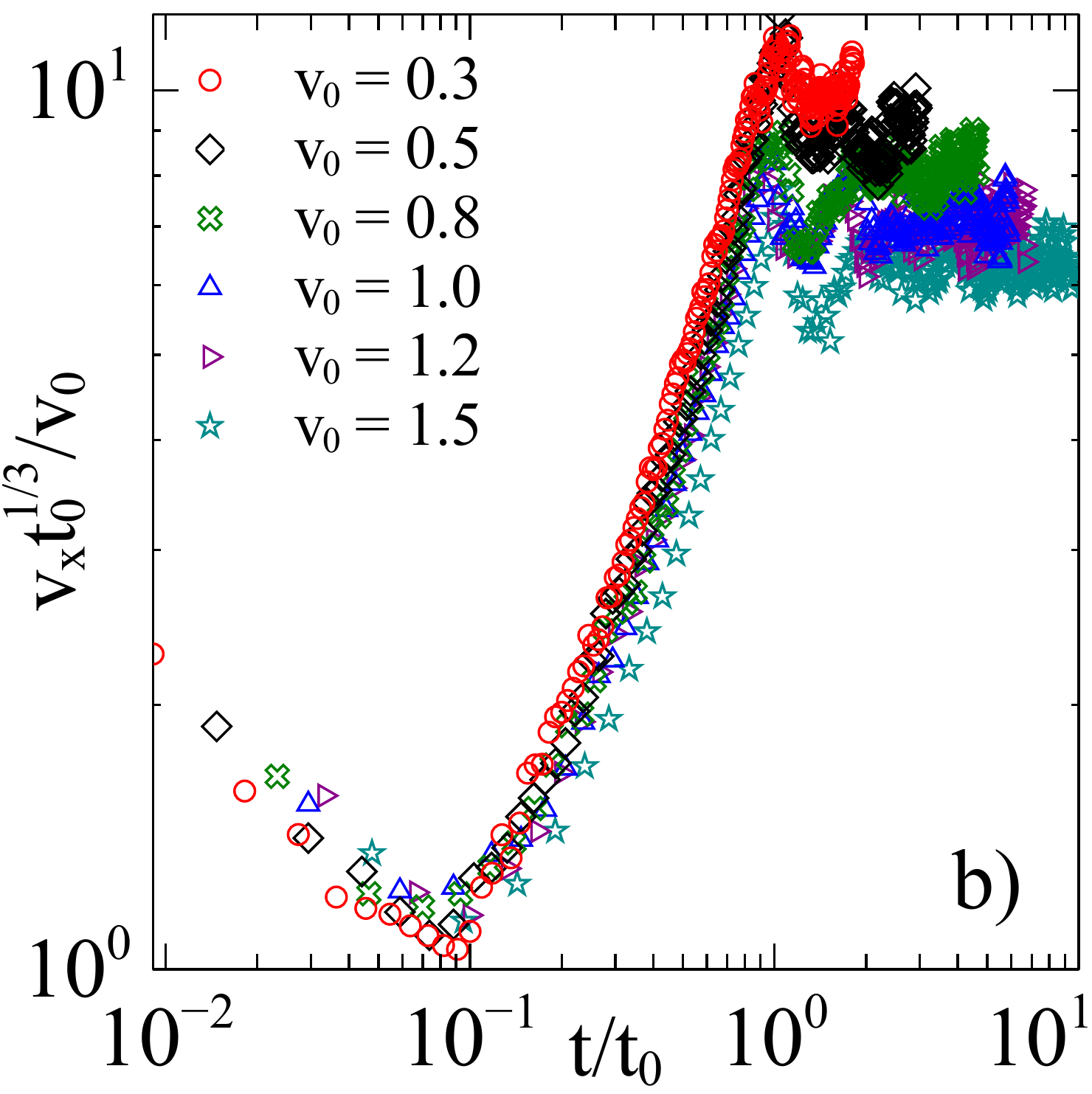}
\caption{(Color online) a) Evolution with time of the ensemble average
of the velocity component $v_x\equiv \langle |v_{x}(t)|\rangle$ for
different values of the shear rate. b) Data collapse of the rescaled
$v_x$ as a function of the rescaled time $t/t_0$.}  \label{fig:vx_t}
\end{figure}

From Figs.~(\ref{fig:PDF2d}) and (\ref{fig:PDF3d}), we notice that the
vortex core structure has a drastic influence on the defect velocity
statistics~\cite{Min96}. This effect is given by a Gaussian tail which
takes over the $v^{-3}$ regimes at very large fluctuations. The vortex
core size is more pronounced in 3D simulations, due to numerical
bounds on higher spatial resolutions. It is easier to vary the aspect
ratio between system size and vortex core size in 2D simulations to
observe the vortex core effect. In CDS simulations, the core size is
fixed by the parameter $A$. We consider two different values of $A$, namely $A=
1.05$ for small cores and $A=2.05$ for larger cores, and averaged over $2000$ random
initial conditions. The system size was reduced to $128^2$ cells. The dependence of the Gaussian-cutoff on the core size is shown in the inset of
Fig.~(\ref{fig:PDF2d}), while the velocity distribution for an intermediate core size, parametrized by $A=1.5$, is plotted in the main graph of Fig.~(\ref{fig:PDF2d}), using $1024^2$ cells and averaging over $48$ initial conditions.

\section{Dislocation statistics}
Next we discuss the statistics of dislocations in anisotropic stripes
during phase ordering and non-relaxational dynamics assisted by a shear
flow.

\subsection{Dynamical scaling regimes}
In the early stages, the evolution of defects is dominated by pairwise
interactions leading to annihilations and a decrease in the density of
defects. We measure the density of defects $\rho_d(t)$ as the ratio
between the effective area occupied by the defects and the total system
area. Alternatively, the number of defects $N$ can be estimated as the
area occupied by the defects divided by the approximate core area of a
defect. In simulations, we choose a large system size of $1024^2$ and
calculate $\rho_d(t)$ as a function of time, starting from a random
initial condition. During the  relaxation dynamics, the defect density
for a very large system is equivalent to the averaged density over many
initial conditions for a smaller system size, and is a smooth function
of time~\cite{Shinozaki93}. As expected in the coarsening regime, the
density obeys a power law in time as $\rho_d(t)\sim \log(t)/t$ like the
density of vortices in Ginzburg-Landau theory~\cite{QianMazenko03}.
This behavior is shown in Fig.~(\ref{fig:rho_t}a) by the data in open
circles and is consistent with formal arguments that the anisotropic
Swift-Hohenberg dynamics can be mapped onto the isotropic
Ginzburg-Landau dynamics~\cite{Pismen99}. In Fig.~(\ref{fig:rho_t}a),
we also plot the density $\rho_d(t)$ as a function of time for various
values of the shear rate $v_0$. We notice that in the late stages, when
long-range hydrodynamic interactions set in, the density of defects
ceases to decrease monotonically, and approaches instead a
statistically steady state.  In this steady state , the defect density
fluctuates in time about a mean value because of the sporadic pair
creations and subsequent annihilations of defect pairs. Averages over
initial conditions would lead to a constant mean density $\rho_0$ in
this steady state, i.e. $\langle\rho_d(t)\rangle_{IC}\rightarrow\rho_0$
as $t\rightarrow\infty$. We also observe that the mean number of
defects $\langle N\rangle$ in the steady state appears to increase with
the applied shear $v_0$, suggesting that the creation rate of defect
pairs depends on $v_0$. Also, the mean square fluctuations $\langle
N^2\rangle-\langle N\rangle^2$ in the number of defects increases
monotonically with $\langle N\rangle$ with significant deviations from
what would be expected in a Poisson process. It appears that the number statistics of
defects behaves in a similar manner to that in defect turbulence~\cite{Huepe04},
although a detailed analysis of this suggestion would be beyond the scope of this paper.

From the data presented in Fig.~(\ref{fig:rho_t}b), we can determine
the cross-over time $t_0$ to the statistical steady state, and we find that
it increases to a first approximation as $t_0\sim v_0^{-1}$, as shown in
the inset of Fig.~(\ref{fig:rho_t}b). It turns out that phase ordering
assisted by hydrodynamic effects slows down the growth of the typical
distance between defects with corrections that follow a $L(t)\sim
t^{1/3}$ law until saturating to the steady state. This implies that
the steady state defect density can be estimated as $\rho_0\sim
L(t_0)^{-2}\sim t_0^{-2/3}$. Equivalently, the mean density in the
steady state increases with the applied shear as $\rho_0 \sim
v_0^{2/3}$. As a first step to see whether there is any data collapse
associated with this scaling behavior, we plot the rescaled
$\rho/\rho_0$ versus the rescaled $t/t_0$ as shown in
Fig.~(\ref{fig:rho_t}b).

\begin{figure}[t]
\includegraphics[width=0.46\columnwidth]{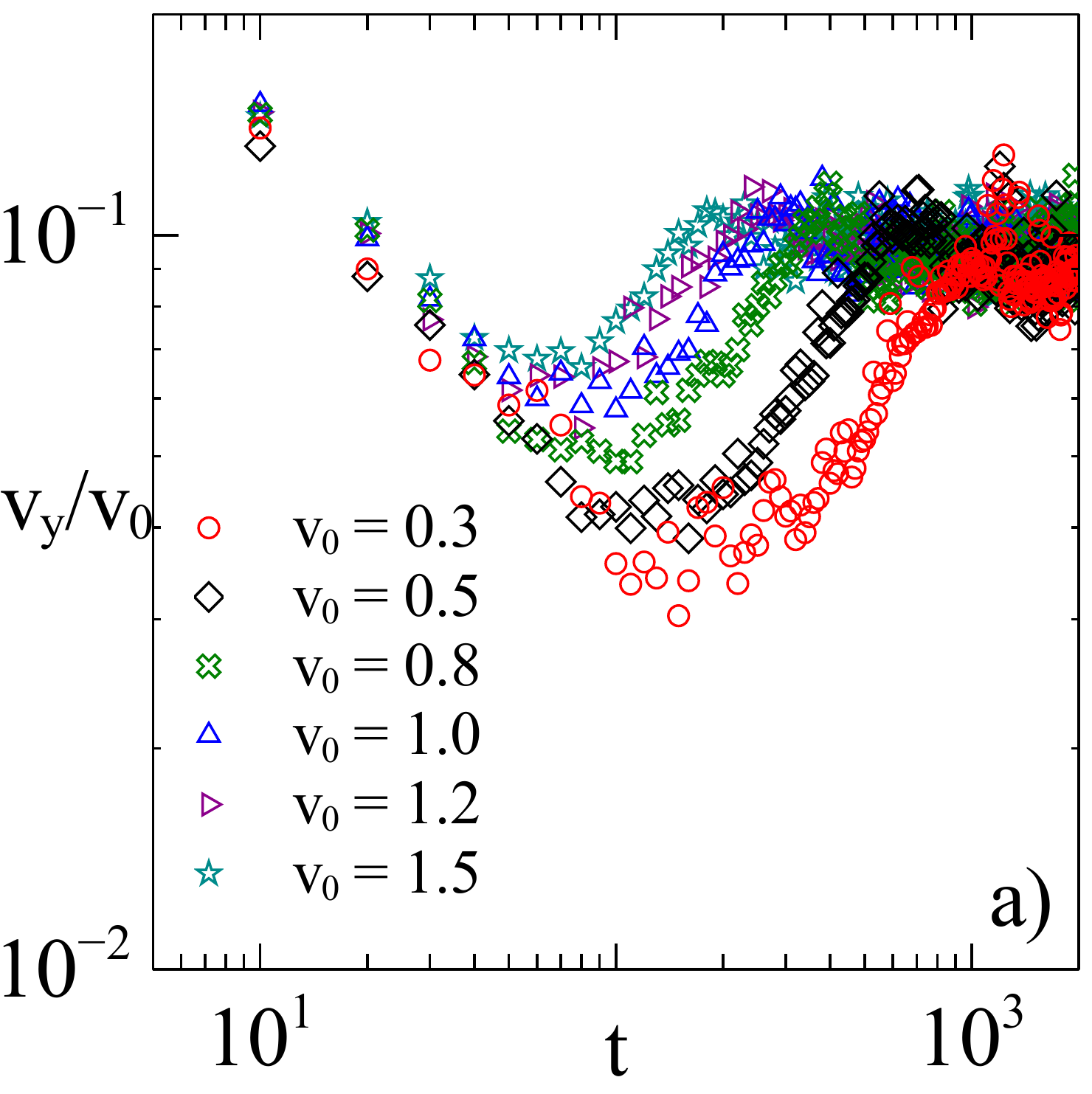}
\includegraphics[width=0.46\columnwidth]{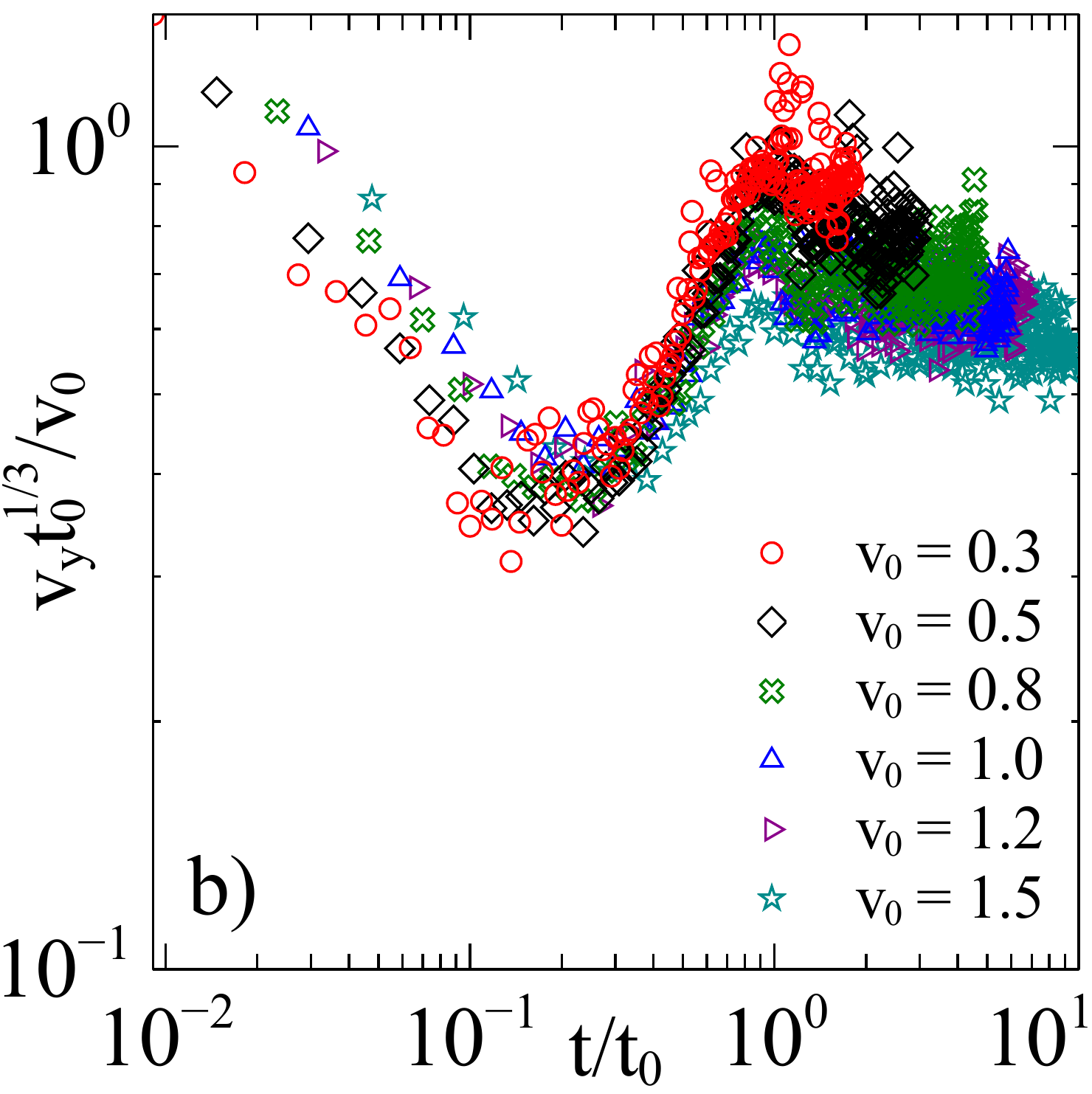}
\caption{(Color online) a) Evolution with time of the ensemble average
of the velocity component $v_y\equiv \langle |v_{y}(t)|\rangle$ for
different values of the shear rate. b) Rescaled $v_y$ plotted as a
function of the rescaled time $t/t_0$.}  \label{fig:vy_t}
\end{figure}

Because the normal stripes are aligned along the $y$-direction, the
climb and glide motions of the dislocations correspond respectively to
the vertical and horizontal velocities. The motion is anisotropic and
dominated by the transverse (gliding) dynamics of dislocations. We
observe that climb motion typically occurs when two dislocations of
opposite topological charge approach each other to annihilate,
otherwise dislocations would move by gliding across the stripes.
Nevertheless, when the two motions are driven by the local pairwise
interactions between dislocations, they exhibit similar
characteristics. In the early states, where the dynamics is controlled
mainly by the phase ordering and annihilations of dislocations, the
ensemble average of the net velocities (absolute values), $\langle
|v_x|\rangle(t)$ and $\langle |v_y|\rangle(t)$, scale with time as
$t^{-1/2}$ until hydrodynamic effects set in and the defects are
accelerating until they cross over at $t_0$ to a statistically
stationary state. This behavior is shown in Fig.~(\ref{fig:vx_t}a) for
$\langle|v_x|\rangle$ and in Fig.~(\ref{fig:vy_t}a) for
$\langle|v_y|\rangle$ corresponding to different values of the imposed
shear velocity $v_0$. In the later stages, hydrodynamic effects become
important and there is a transient regime where the mean velocities
increase almost linearly with time until $t_0$ after which a
statistically stationary state is reached. Since, we run large scale
simulations for a given realization without averaging over initial
conditions, the ensemble-averaged velocities, $\langle|v_x|\rangle(t)$
and $\langle|v_y|\rangle(t)$, correspond to time series in the
statistical steady state. Averaging these fluctuations over many
initial conditions would smooth out the late-stage time dependence to a
constant value.

The mean square fluctuations in the steady state are an increasing
function of the applied shear. We observe that the mean value of climb
velocity is approximately an order of magnitude smaller than the
typical velocity of gliding. A rescaling of velocity components as a
function of the rescaled time in the units of $t_0$ is presented in
Figs.~(\ref{fig:vx_t}b) and (\ref{fig:vy_t}b), which however gives a
poor data collapse. Improving it turned out to be a challenging task,
one of the reasons being that there is an additional characteristic
timescale given by the cross-over from the relaxation dynamics to the
transient  period of acceleration prior to the steady state. It may be
that this timescale also plays a role in the scaling function, but we
have not succeeded in including it to our satisfaction. This is an
unresolved issue that deserves a separate detailed study.

The statistically stationary regime is characterized by fluctuations in
the density of defects and their velocities around a mean value that
depends on the imposed shear flow. Fluctuations in the defect density
are attributed to sudden nucleation of dislocation pairs and their
subsequent annihilation, either with the same pair member, or with
dislocations from another pair.

\begin{figure}[t]
\includegraphics[width=0.9\columnwidth]{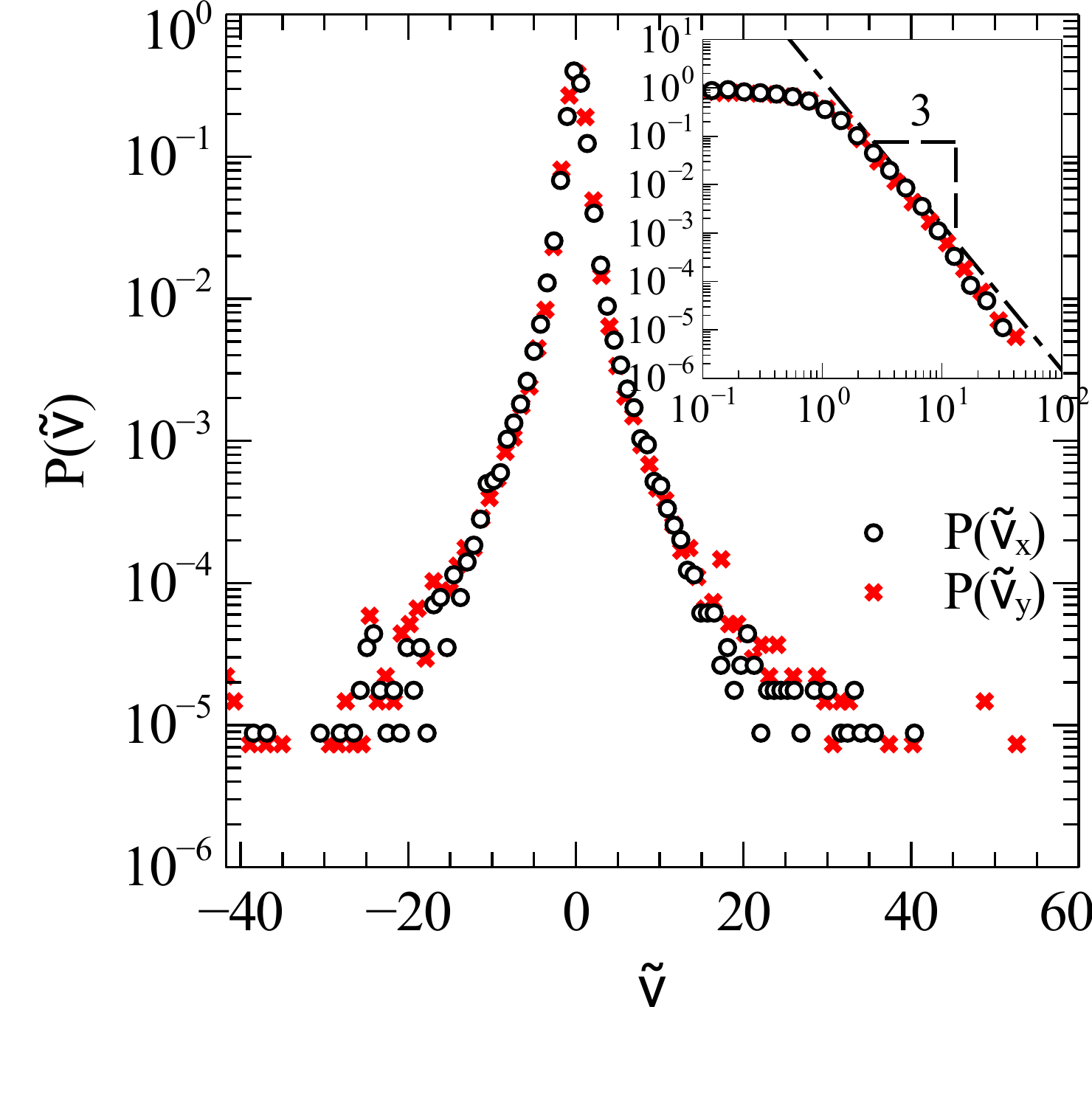}
\caption{(Color online) The time independent scaling function of the
probability distribution corresponding to climb (crosses), respectively
glide (open circles) velocities during the phase ordering in the
absense of external field. The rescaled velocity variable is given as
$\tilde v_j\equiv v_j/\langle |v_j|\rangle$ for $j=x,y$. In the inset figure it is shown the log-log plot of the PDF calculated with logarithmic binning.} \label{fig:PDF_SH}
\end{figure}

Physically, the nucleation mechanism is related to the phase shifts
induced by the transverse shear deformations. The reason for this is
that the action of an external shear flow is cumulative in the phase
$\theta$ of the complex envelope field $\psi = |\psi|e^{i\theta}$ and
its gradients $\nabla\theta$, similar to the effect of the self-induced
mean flow~\cite{Pismen92}. The shear flow advects the underlying stripe
pattern together with its defects. This motion induces small
undulations along the stripes which build up stresses and create
distortions in the pattern. These distortions, which are localized into
transverse~\lq shear bands\rq~, grow with time up to the point where
they locally tear apart the stripes releasing pairs of defects. The
shear flow affects both the isolated motion of defects and, more
importantly, the interaction between defects. The effect of the
large-scale flow on the defect interactions becomes stronger where the
defect motion is slower~\cite{Pismen92}.

\subsection{Velocity statistics}

In the absence of shear flow, the motion of dislocations is symmetric.
Although the mean velocity in absolute value is non-zero and related to
the mutual interaction forces, the velocity of dislocations averages
out to zero. This is equivalent to the symmetric probability
distributions of the velocity fluctuations as shown in
Fig.~(\ref{fig:PDF_SH}). The actual distribution is time dependent due
to quench dynamics by annihilations. However, using the dynamical
scaling behavior of the probability distribution we can remove the time
dependence by effectively rescaling the dislocation velocity by the
ensemble average of the absolute velocity at a given time, i.e. $\tilde
v_i \equiv v_i/\langle |v_i(t)|\rangle$, with $i=x,y$ for the two
components. The distribution of these rescaled velocities corresponds
to the scaling function of the time dependent probability distribution
as discussed previously in the context of vortex dynamics. From
Fig.~(\ref{fig:PDF_SH}), we notice that the probability distributions
of the climb and glide velocities retain a similar form that is
characterized by a long tail with a $-3$ power law as in the
relaxational dynamics of vortices. This is consistent with previous
numerical studies of the velocity statistics of defects in the
anisotropic Swift-Hohenberg dynamics from Ref.~\cite{Mazenko06}.

At a finite shear rate and in the late stage of statistically
stationary regime, the motion of dislocations is influenced by the
imposed flow. Dislocations of opposite topological charges tend to move
in opposite directions, with an asymmetry in their mean transverse
motion that is related to the shear rate, i.e. $\dot\gamma\sim(\langle
v^+_x\rangle-\langle v^-_x\rangle)$, while the climb motion remains
almost symmetric. The velocity probability distribution for positive,
respectively negative dislocations becomes the same when we rescale
their corresponding absolute velocities as $\tilde v^s\equiv
|v^s|/\langle |v^s|\rangle$, where $s= \pm$, so that $P^+(\tilde v^+)=
P^-(\tilde v^-)$.

\begin{figure}[t]
\includegraphics[width=0.9\columnwidth]{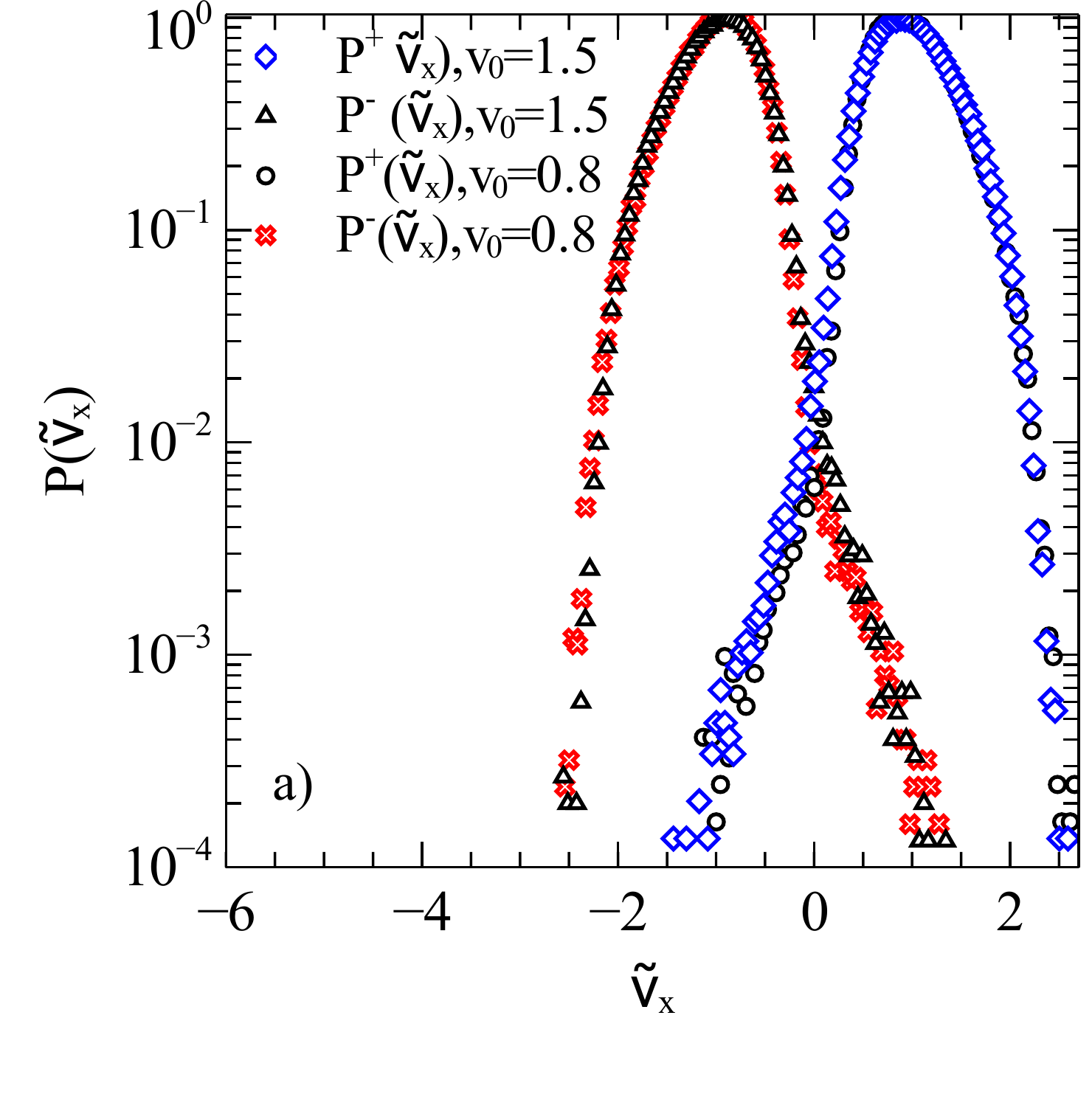}
\includegraphics[width=0.9\columnwidth]{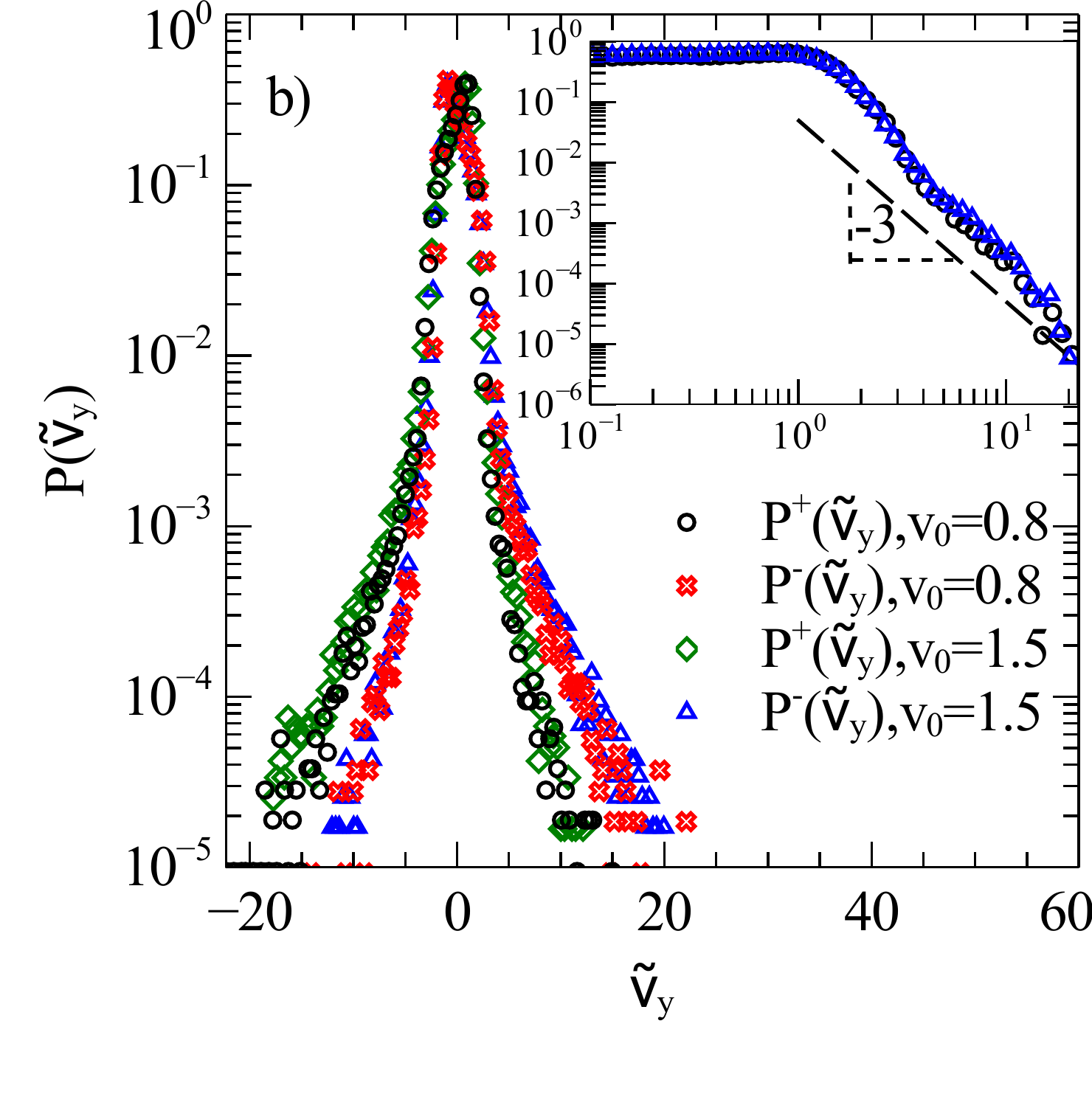}
\caption{(Color online) (a) PDF of the gliding velocity for positive (open diamonds and circles) and negative (open triangles and crosses) dislocations in the statistically stationary state at a different shear rates. (b) Probability distribution function of the climb velocity for positive (open circles and diamonds) and negative (open crosses and triangles) dislocations in the statistically stationary state. In both cases, the rescaled velocity variable is $\tilde v_j\equiv v_j/\langle |v_j|\rangle$ for $j=x,y$. Inset figure shows the log-log plot of the tail PDF with logarithmic binning.}
\label{fig:PDF_SH_shear}
\end{figure}

In Fig.~(\ref{fig:PDF_SH_shear}), we plot the probability distribution
functions of the rescaled gliding velocities and climbing velocities in
the steady state regime. For the transverse motion, the small scale
velocity fluctuations are normally distributed around the mean flow.
Large fluctuations against the flow are due to events where pairs of
dislocations of opposite charge, gliding opposite to their drift flow,
are attracting and annihilating. These events contribute to the long
left tail for $P^+(v_x)$ and right tail for $P^-(v_x)$. The statistics
of small climb velocities are also influenced by the imposed shear flow
and given by the slight asymmetry in the $P^s(v_y)$. However, the large
fluctuations in the longitudinal motion are due to pair interactions
prior to annihilations. These fluctuations are captured by the long
tails in the $P(v_y)$ distribution and they seem to follow the inverse
cubic law, but with less accuracy than in the relaxational case.
Typically, a nucleation event leads to a burst in the local density of
dislocations which will annihilate subsequently by the fast climbing
motions. However, these large velocity events occur on a longer
timescale than during relaxational dynamics, so that the system needs
to be followed longer in the steady state. This is computationally
challenging, because of the prior transient acceleration period whose
length increases as the driving force is decreased.

\section{Conclusions:} In summary, our numerical simulations suggest
that the velocity statistics of codimensional-two defects exhibits a
dynamical scale invariance with a scaling function that has a
universal inverse cubic tail when the defect dynamics is driven by
mutual pair interactions leading to annihilations. This is valid both
for point defects in 2D and defect filaments in 3D during phase
ordering kinetics. Finite size effects of the defect core introduce a
Gaussian cut-off to the $v^{-3}$ scaling. A similar statistical
behavior is observed in the velocity of dislocations in anisotropic
stripe patterns during phase ordering. Although the motion is highly
anisotropic and dominated by the gliding of dislocations, the
distributions of the climb and glide velocity fluctuations exhibit the
same algebraic tail when the motion is driven by the local interactions
between dislocations. During non-relaxational dynamics assisted by an
external transverse shear flow, small velocity fluctuations are
influenced by the mean flow, whereas the asymptotically large
fluctuations are still due to pairwise interactions. In
statistically stationary dynamics,  anisotropic defect motion is
manifested also in anisotropic statistics of the glide and climb
velocities. In this study, we have neglected the effect of a
self-induced mean flow in the defect dynamics. This effect is however
important to capture the spatio-temporal chaotic dynamics as seen from
experiments. It is thus interesting to study in future investigations
the statistical properties of interacting defects when the combined
effect of a large scale flow and a self-induced mean flow is taken into
account.

{\it Acknowledgements:} This work is partially supported from the
Center for Physics of Geological Processes at UiO and by US National
Science Foundation Grant DMR 1044901.

\appendix*

\section{Calculation of derivatives in cell dynamical systems models}
We solve the time dependent real Ginzburg-Landau dynamics from
Eq.~(\ref{eq:TDGL}) using the cell dynamical systems approach
(CDS)~\cite{Oono1988,Mondello1990,Mondello1992}. Numerical efficiency
becomes particularly important in 3D simulations and this method is
tailored to that. The complex variable $\psi(\bm r,t)$ is replaced by a
$\psi_{i,j}^{(n)}$ (or $\psi_{i,j,k}^{(n)}$ in 3D) defined on a square
lattice of size $N\times N$ (a cube of size $N\times N\times N$ in 3D)
at time $n$. The idea of the CDS method is to construct a discrete set
of maps for each lattice cell such that the flow properties of the
continuous dynamical system are preserved. A cell dynamics is defined
by two steps: a local update
\begin{equation}
\tilde\psi^{(n+1)} = \frac{A\psi^{(n)}}{\sqrt{1+\psi^{2(n)}(A^2-1)}},
\end{equation}
where $A>1$ is a parameter that determines the global rate of
convergence to the fixed points of the local double-well potential, and
a global update taking into account the interactions between
neighbouring cells
\begin{equation}
\psi^{(n+1)} = \tilde\psi^{(n+1)}+C\nabla^2\tilde\psi^{(n+1)},
\end{equation}
where $C$ is a constant proportional to the phenomenological diffusion
constant. The isotropy of the order parameter being simulated naturally
mandates the isotropy of the difference operators used to implement the
coupled maps (see Tomita \cite{Tomita91}). Oono and
Puri~\cite{Oono1988} chose a 9 point stencil to implement a
``Laplacian'' operator, which is highly isotropic for a 2D square
lattice \cite{Teixeira1997}. This stencil reads
%
\begin{eqnarray}
\nabla^2\psi\equiv \frac{3}{dx^2}\left(\frac{1}{6}\sum_{\mathrm{NN}}
\psi + \frac{1}{12}\sum_{\mathrm{NNN}}\psi - \psi\right),
\end{eqnarray}
where $NN$ stands for the nearest neighbours in the discretized lattice
and $NNN$ are the next-to-nearest neighbours for each node in the
lattice.

Considering the same isotropy requirements, the discretization of the
3D Laplace operator reads as~\cite{Mondello1992}
\begin{eqnarray}
\nabla^2f \equiv \frac{3}{dx^2}\left(\frac{1}{9}\sum_{\mathrm{NN}} \psi + \frac{1}{36}\sum_{\mathrm{NNN}}\psi - \psi\right).
\end{eqnarray}
%

Since the calculation of the position and velocities of defects
involves first order spatial derivatives of the $\psi$ field, an
isotropic discretization of the gradients is important in order to
reduce the underlying lattice anisotropic effects. We use the isotropic
version of the gradients both in 2D and 3D. Following the idea that
these operators have to be accurately represented in Fourier space
\cite{Teixeira1997}, we use the following stencil for 2D
\begin{eqnarray}
\nabla_x\psi &\equiv& \frac{1}{8dx}\left(\psi_{i+1,j+1} + 2\psi_{i,j+1} - \psi_{i-1,j+1}\right. + \nonumber \\
&+& \left. \psi_{i+1,j-1}-2\psi_{i,j-1}-\psi_{i-1,j-1}\right) ,
\end{eqnarray}
where $i$ and $j$ are the lattice indices for the $x$ and $y$
directions, respectively. Swapping indices we can obtain the
corresponding expression for $\nabla_y \psi$. We note that this
expression looks very similar to a 4 point first derivative in a 2D
square lattice \cite{Davis1970}. For the gradients in 3D, the stencil
reads
\begin{widetext}
\begin{eqnarray}
\nabla_x\psi &=& \frac{1}{8dx} \left(\psi_{i+1,j+1,k}- \psi_{i-1,j+1,k}+ \psi_{i+1,j-1,k} - \psi_{i-1,j-1,k}+\psi_{i+1,j,k+1}-\psi_{i-1,j,k+1}+\psi_{i+1,j,k-1} - \psi_{i-1,j,k-1}\right) + \nonumber \\
&+& \frac{1}{4dx} \left(\psi_{i,j+1,k}-\psi_{i,j-1,k}+\psi_{i,j,k+1}-\psi_{i,j,k-1}\right) ,
\end{eqnarray}
with the corresponding index swap to obtain $\nabla_y\psi$ and $\nabla_z\psi$.
\end{widetext}

\bibliographystyle{apsrev4-1}
\bibliography{references}
\end{document}